\documentclass[aps,prb,twocolumn,groupedaddress,showpacs,floatfix]{revtex4}
\usepackage{graphics,graphicx}
\usepackage{color}
\usepackage{wrapfig}
\usepackage{feynmf}
\usepackage{amssymb,amsmath}
\usepackage{bookmath}
\usepackage{bm}

\begin{document}
\unitlength = 1mm
\title{Interplay between magnetism and superconductivity in Fe-pnictides}
\author{A.~B.~Vorontsov, M.~G.~Vavilov, and A.~V.~Chubukov}
\affiliation{Department of Physics,
             University of Wisconsin, Madison, Wisconsin 53706, USA}

\date{December 12, 2008}
\pacs{74.25.Dw,74.25.Ha}

\begin{abstract}
We consider phase transitions and potential co-existence of
spin-density-wave (SDW) magnetic order and extended $s$-wave ($s^+$)
superconducting order within a two-band itinerant model of iron
pnictides, in which SDW magnetism and $s^+$ superconductivity are
competing orders. We show that depending on parameters, the
transition between these two states is either first order, or
involves an intermediate phase in which the two orders co-exist. 
We demonstrate that such co-existence is possible when SDW order is
incommensurate.
\end{abstract}
\maketitle

{\it Introduction.} Iron-based pnictide superconductors -- oxygen
containing 1111 materials RFeAsO (R = La, Nd, Sm) and oxygen free
122 materials AFe$_2$As$_2$ (A = Ba, Sr, Ca) are at the center of
experimental and theoretical activities at the moment because of
high potential for applications and for the discovery of new
mechanisms of superconductivity. Most of parent compounds of
Fe-pnictides are magnetically ordered. Upon doping, magnetism
disappears and superconductivity emerges, but the nature of this
transition remains unclear. Some experiments on fluoride-doped 1111
materials indicate that the transition is
first-order,\cite{luetkens08} some show behavior more consistent
with a quantum-critical point separating magnetic and
superconducting (SC) states~\cite{zhao08}, and some show a
co-existence of magnetism and SC in both classes of
 materials.\cite{chen08,kamihara08,liu08}

It is by now rather firmly established that Fe-pnictides are metals
in a paramagnetic phase for all dopings, with two sets of (almost)
doubly degenerate Fermi surface (FS) pockets -- a hole pocket
centered at $(0,0)$ and an electron pocket centered at $\vpi
=(\pi,\pi)$ in the folded Brillouin zone. To a good approximation,
hole and electron FS are circular and at zero doping have nearly
identical sizes.~\cite{singh08,cliu08a,cliu08b,sebastian08,kondo08,ding08} 
Like in chromium (Ref.~\onlinecite{kulikov84}),  
this nesting geometry is favorable to a
spin-density-wave (SDW)  ordering at $\vpi$  as the corresponding
susceptibility logarithmically diverges at $T=0$,\cite{kulikov84} and a
small repulsive interaction in the particle-hole channel at momentum
transfer $\vpi$ already gives rise to an SDW instability at $T =
T_s$. If the interaction is attractive in a SC pairing channel, then
the SC pairing susceptibility also diverges logarithmically at $T=0$
and the system becomes a SC at $T = T_c$, unless magnetism
interferes.

Previous  studies of an itinerant model of small electron and hole FS have
found that the \emph{same} interaction, inter-band Josephson-type pair hopping,
gives rise to an SDW order and to superconductivity with extended $s$-wave
($s^+$) symmetry of the SC order parameter
($\Delta (k) \propto \cos k_x + \cos k_y$
in the folded Brillouin zone),\cite{chubukov08,d_h_lee}
leading to competition between the two orders.
The full interactions in SDW and $s^+$ channels also involve inter-band
forward scattering and intra-band Hubbard interaction, respectively, and
the full interaction is  larger in the SDW channel.~\cite{chubukov08}
Then at zero doping, which we associate with near-perfect nesting,
the highest instability temperature is that of
an SDW state. At a nonzero doping $x$, nesting is destroyed (either
hole or electron pocket gets relatively larger), and SDW
susceptibility no longer diverges. Magnetic $T_s (x)$ then goes down
with doping and above a particular value of $x$, the first
instability upon cooling is into $s^+$ SC state. The superconducting
state is only weakly affected by doping.

The goal of the present work is to understand how the system evolves from an
SDW antiferromagnet to an $s^+$ superconductor. For this we derive and solve a
set of coupled non-linear BCS-type equations for SC and SDW order
parameters. We assume that the interactions in the two channels are comparable
in strength and that $T_c \lesssim T_s$, where  $T_s$ is
the transition temperature at zero doping, $T_s=T_s(x=0)$.

\begin{figure}[t]
\centerline{\includegraphics[width=\linewidth]{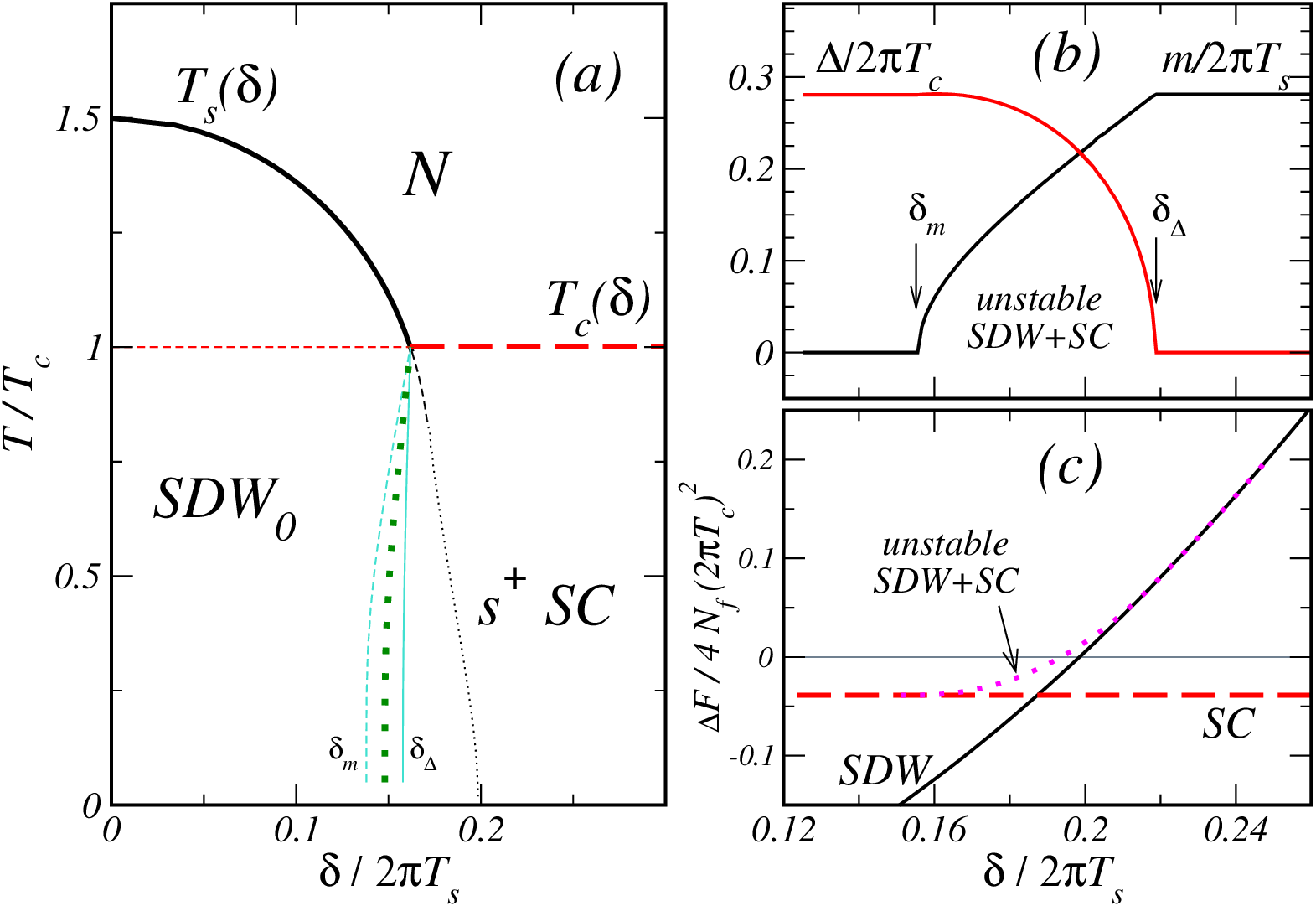}}
\caption{(color online) (a) Phase diagram for $T_s/T_c=1.5$
as a function of $\delta$ controlled by doping.
Thick solid and dashed lines are second-order SDW and SC transitions at
 $T_s(\delta)$ and $T_c$, and dotted line - first order transition
between commensurate SDW$_0$ and $s^+$ SC.
Thin lines - physically unaccessible transitions.
The pure magnetic $T_s(\delta)$ line follows the
curve of `paramagnetically limited superconductivity'.
The superconducting $T_c$ is independent of
 $\delta$ in our model.
Light lines denoted $\delta_\Delta$ and $\delta_m$
are instability lines of SC and SDW states.
(b) SDW and SC order parameters, and
(c) free energy for SDW, SC and unstable mixed phases at $T/T_c=0.1$.
}
\label{fig:comm}
\end{figure}

\begin{figure}[t]
\centerline{\includegraphics[width=\linewidth]{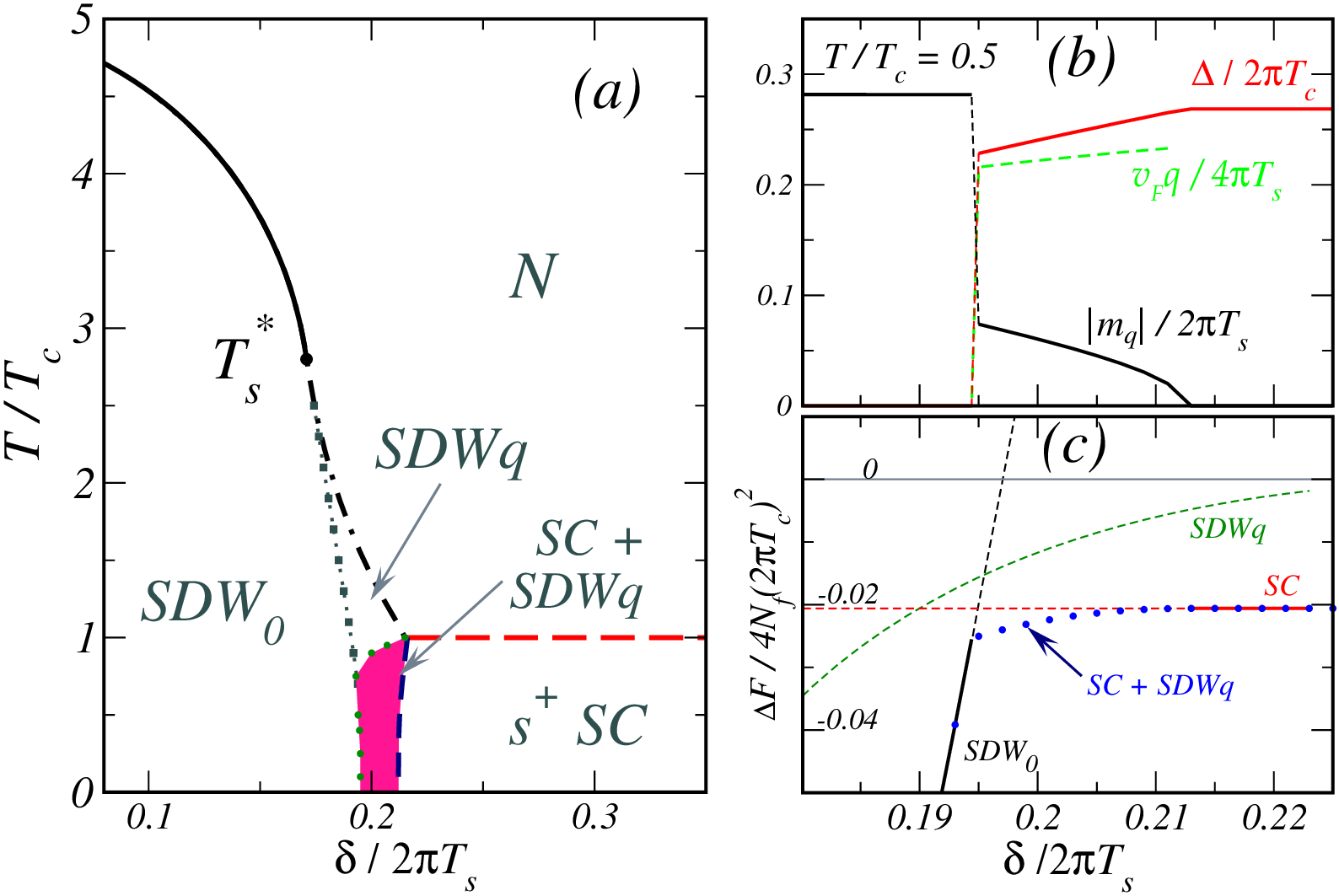}}
\caption{(color online)
Same as in Fig. \protect\ref{fig:comm}, but for $T_s/T_c =5$.
An incommensurate SDW order appears below $T_s (\delta)$ once
it becomes smaller than $T^*_s=0.56 T_s > T_c$.
Below $T_c$, a new mixed phase appears in which incommensurate SDW$_q$ order
co-exists with SC. At small $T$, there is no SDW$_q$ state without
superconductivity. The transitions into the mixed state are second order
from a SC state and first order from a commensurate SDW state.
The free energy now shows that near $\delta/2\pi T_s =0.2$
a mixed state has lower energy than the two pure states.
}
\label{fig:incomm5}
\end{figure}

\begin{figure}[t]
\centerline{\includegraphics[width=0.9\linewidth]{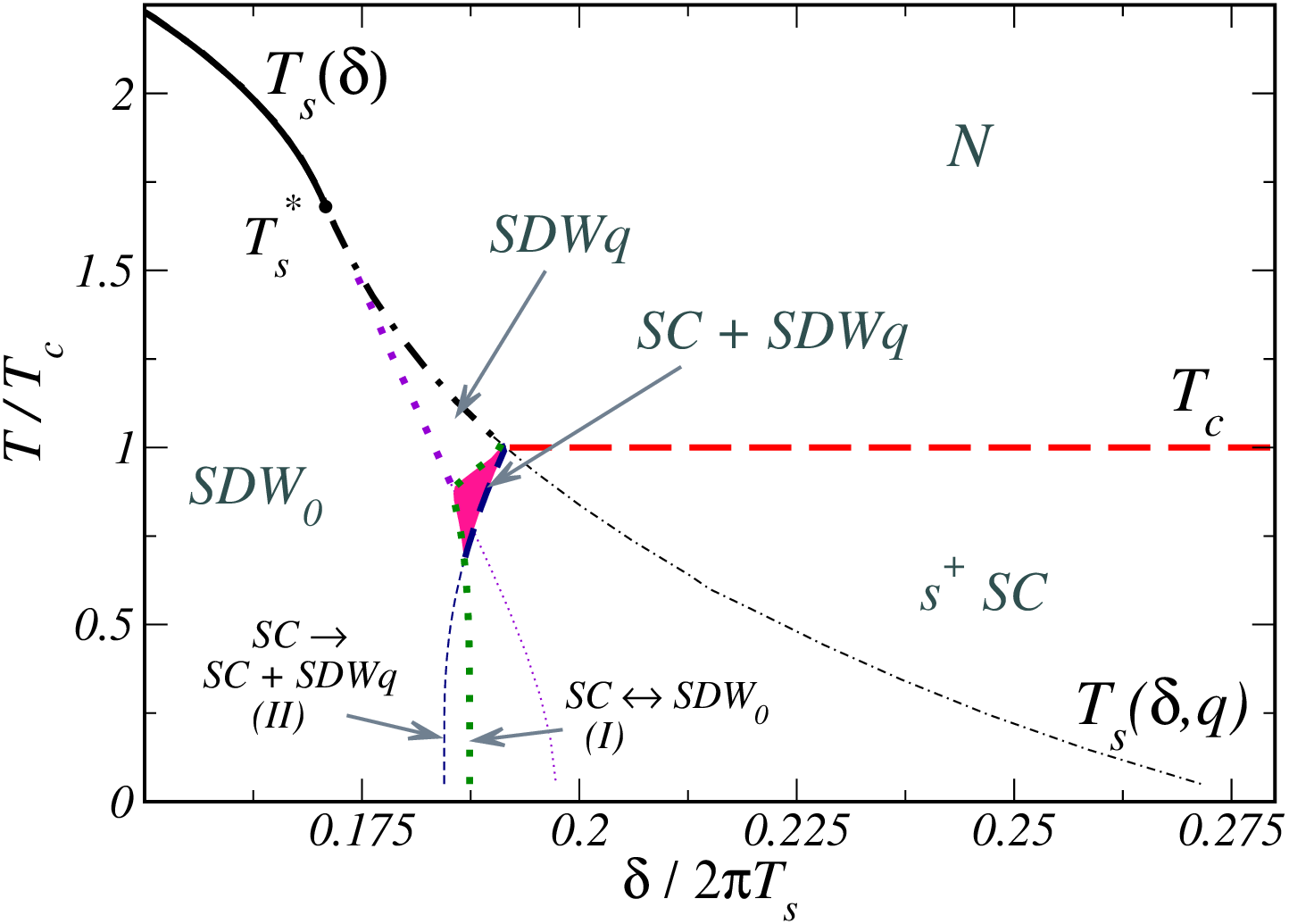}}
\caption{(color online)
Same as in previous two figures, but for intermediate $T_s/T_c =3$. Mixed
phase appears only in a tiny region near the point where $T_c$ and
$T_s (\delta,q)$ cross. At smaller $T$, the system still displays a first order
transition (dotted line) between a commensurate SDW$_0$ state and a SC state.
}
\label{fig:incomm3}
\end{figure}

We report two results. First, when $T_s/T_c$ is close to unity, the system
displays second order SDW and SC transitions at $T_s(x)$ and $T_c$, whichever
is larger. The SDW state is commensurate, with momentum $\vpi$.  At smaller $T$,
the transition between SDW and SC upon changing $x$ is first order,
and there is no stable co-existence region (Fig.~\ref{fig:comm}).
This is similar to the phase diagram
reported for LaFeAsO$_{1-x}$F$_x$ in Ref.\onlinecite{luetkens08}.
Second, when $T_s/T_c$ gets larger,
SDW order becomes incommensurate with momentum $\vQ = \vpi +\vq$
below some $T^*_s = 0.56 T_s > T_c$ (an SDW$_q$ phase.\cite{rice70,cvetkovic08})
We argue that in this situation  SDW$_q$ and SC states co-exist.
The co-existence region is initially confined to a small region below $T_c$,
while at lower $T$ the system still displays a first order transition
between a commensurate SDW and SC states. As the ratio $T_s/T_c$ increases,
the co-existence region
extends down to lower $T$  and eventually
reaches $T=0$ (Figs.~\ref{fig:incomm5},\ref{fig:incomm3}).

An incommensurate SDW$_q$ state at finite dopings has been studied
in connection with theoretical models for chromium and its alloys
by Rice~\cite{rice70} and others,\cite{kulikov84} and in connection to pnictides by
Cvetkovic and Tesanovic.\cite{cvetkovic08} Such state
is a magnetic analog of Fulde-Ferrell-Larkin-Ovchinnikov (FFLO) state,\cite{FFLO} for
which doping plays the same role as a magnetic field
in a paramagnetically limited superconductor.\cite{rice70}
We found that at $T<T_c$ SDW$_q$ phase exists {\it only}
in combination with $s^+$ superconductivity.

{\it Model and equations.} We neglect double degeneracy of hole and electron
states, which does not seem to be essential for the pnictides,\cite{d_h_lee,korshunov08}
and consider a weak-coupling model with two families of fermions,
near one hole and one electron FSs of small and
near-equal sizes. The free electron part of the Hamiltonian is
$
\cH_0 = \sum_{\vk}\left( \xi_c(\vk) c^\dag_{\vk\alpha}
c_{\vk\alpha} +\xi_f(\vk) f^\dag_{\vk\alpha} f_{\vk\alpha} \right)$,
where operators  $c_{\vk\alpha}$ and $f_{\vk\alpha}$ describe fermions
near $(0,0)$ and $(\pi,\pi)$, respectively (the momentum $\vk$ in
$f_{\vk\alpha}$ is a deviation from $\vpi$). The dispersions
$\xi^{f,c}(\vk) = \pm \xi_\vk + \delta$, where $\xi_\vk = v_F
(k-k_F)$, and $\delta$ measures a deviation from a prefect nesting
and
may be tuned by doping($x$) or pressure.
The four-fermion part contains interactions in SDW and SC channels and in
 mean-field (BCS) approximation reduces to the effective quadratic form
$\cH = \onehalf \sum_{\vk\alpha\beta} \;
\overline{\Psi}_{\vk\alpha} \, \hat\cH \, \Psi_{\vk\beta}$,
with
$\overline{\Psi}_{\vk\alpha} =
(c^\dag_{\vk\alpha}, c_{-\vk\alpha}, f^\dag_{\vk+\vq\alpha}, f_{-\vk-\vq\alpha})$,
($\Psi$- is a conjugated column):
\begin{widetext}
\be
\hat\cH =
\left( \begin{array}{c|c}
\begin{array}{cc} \xi^c(\vk) &   \Delta^c \; i \sigma^y_{\alpha\beta} \\
-\Delta^{*c} \; i\sigma^y_{\alpha\beta} & -\xi^c(-\vk) \end{array}
&
\begin{array}{cc} m_\vq \; \sigma^z_{\alpha\beta} & 0 \\
0 & - m_\vq \;  \sigma^z_{\alpha\beta} \end{array}
\\ \hline
\begin{array}{cc} m^*_\vq \; \sigma^z_{\alpha\beta} & 0 \\
0 & - m^*_\vq \; \sigma^z_{\alpha\beta} \end{array}
&
\begin{array}{cc} \xi^f(\vk+\vq) &  \Delta^f \; i\sigma^y_{\alpha\beta} \\
-\Delta^{*f} \; i\sigma^y_{\alpha\beta} & -\xi^f(-\vk-\vq) \end{array}
\end{array} \right)
\,.
\ee
\end{widetext}
The two diagonal blocks of the matrix $\hat\cH$
include  the $s^+$ SC order parameter $\Delta^c=-\Delta^f=\Delta$
for two FS pockets and two off-diagonal blocks
contain SDW parameter $m_\vq$; 
$\xi^f(\vk+\vq) = \xi_\vk + \delta + \vv_F \vq$ for $q \ll k_F$.
The values of $m_{\bm q}$ and $\Delta$ are
determined by conventional self-consistency equations
\begin{subequations}
\bea
m_\vq  =  V^{sdw} \sum_{\vk} \sigma^z_{\alpha\beta}
\langle f^\dag_{\vk+\vq \alpha} c_{\vk \beta} \rangle
\,, \quad \\
\Delta  =  V^{sc} \sum_{\vk} (-i\sigma^y)_{\alpha\beta}
\langle c_{-\vk \alpha} c_{\vk \beta} \rangle\,,
\eea
\label{eq:sceqs}
\end{subequations}
where the sums are confined to only $(0,0)$ FS pocket,
and  $V^{sdw}$ and $V^{sc}$ are the  couplings in the
particle-hole SDW channel and
in the particle-particle SC $s^+$ channel.\cite{chubukov08}
Taken alone,
$V^{sc}$ leads to an $s^+$ SC state with critical temperature $T_c$,
while $V^{sdw}$ leads to an SDW state with transition temperature $T_s$ at
$\delta=0$.
The SDW order yields a real magnetization
$M(R) \sim \cos\vQ\vR$ at wave vector $\vQ = \vpi +\vq$.
The couplings $V^{sdw}$ and $V^{sc}$
 undergo logarithmic renormalizations from fermions with energies between
$\epsilon_F$ and much larger bandwidth $W$ and flow to the same
value when $W/\epsilon_F \rightarrow \infty$
(Ref.~\onlinecite{chubukov08}).
For any finite $W/\epsilon_F$, $V^{sdw}$ is the largest of the two.

The correlators in Eqs.~(\ref{eq:sceqs}) are related to
components of the Green's function $\whG(\vk,\tau)_{\alpha\beta} =
-\langle T_\tau \Psi(\tau)_{\vk\alpha} \overline\Psi(0)_{\vk\beta}
\rangle$, defined as the inverse of $\whG^{-1} = i\vare_n -
\hat\cH$, where $\vare_n=\pi T(2n+1)$ are the  Matsubara
frequencies. The Green's functions in Eqs.~(\ref{eq:sceqs}) can be
explicitly integrated  over $\xi_- = [\xi^f (\vk+\vq) - \xi^c
(\vk)]/2 = \xi_\vk + \onehalf \vv_F \vq$. Removing the coupling
constants $2N_f |V^{sc}|$ and $2N_f |V^{sdw}|$ ($N_f$ is the density
of states at the Fermi surface per spin) and the upper cutoffs of the
frequency sums in favor of the transition temperatures $T_{c,s}$, we obtain from
Eqs.~(\ref{eq:sceqs})
\begin{subequations}\label{eq:scboth}
\begin{equation}
\ln \frac{T}{T_c} = 2 \pi  T \sum_{n >0 }
\Re \left\langle \frac{ (E_n+i\delta_\vq)/E_n }{\sqrt{(E_n+i\delta_\vq)^2+m_q^2} }
-\frac{1}{|\vare_n|} \right\rangle
\label{eq:scSCt}
\end{equation}
and
\begin{equation}
\ln \frac{T}{T_s} = 2 \pi T \sum_{n > 0}\;
\Re \left\langle \frac{ 1}{ \sqrt{(E_n+i\delta_\vq)^2+m^2_q} }
-\frac{1}{|\vare_n|} \right\rangle.
\label{eq:scSDWt}
\end{equation}
\end{subequations}
where angle brackets denote Fermi surface average,
$E_n =  \sqrt{\vare_n^2 + |\Delta|^2 }$,
$\delta_\vq = [\xi^c (\vk) + \xi^f (\vk+\vq)]/2 = \delta + \onehalf \vv_F \vq$,
and $T_s$ and $T_c$ are
solutions of the linearized equations, respectively for
SDW ($\Delta =\delta=0$) and SC ($m_q=0$).
This system of equations is solved to find all possible uniform SC and
(generally) incommensurate SDW states.

Note that for $\Delta =0$,
Eq. (\ref{eq:scSDWt}) for SDW order
coincides with the gap  equation for
a paramagnetically limited superconductor
with $m_q$ instead of superconducting order parameter,
$\delta$ instead of magnetic field $H$, and incommensurateness
vector $\vq$ instead of the total momentum of a
Cooper pair.\cite{rice70,kulikov84,cvetkovic08}

We will also need the free energy $F(\Delta, m_q)$ for these mean-field order
parameters to pick out the state with  the lowest $F$.
We find $F(\Delta,m_q)$  in two complementary approaches:
by differentiating with respect to interaction parameters,\cite{machida81b} and
 by using Luttinger-Ward functional and extending to a finite $m_q$
the derivation of the condensation energy for a BCS SC.\cite{bardeen_stephen,rainer76}
Both methods yield,
\bea
&&
\hspace*{-1cm}
\lefteqn{
\frac{\Del F(\Delta, m_q)}{4N_f} =
\frac{|\Delta|^2}{2} \ln\frac{T}{T_c} + \frac{ m^2_q}{2} \ln\frac{T}{T_s} -
\pi T \sum_{\vare_m} Re\,
}
\nonumber
\\
&&
\left\langle
\sqrt{(E+i\delta_q)^2+m^2_q} - |\vare_m| - \frac{|\Delta|^2}{2|\vare_m|}
- \frac{m^2_q}{2|\vare_m|}
\right\rangle,
\eea
where $\Del F (\Delta, m_q) = F(\Delta, m_q) - F(0,0)$. We solve
self-consistency equations (\ref{eq:scboth})
for $\Delta$ and $m_q$ at finite $\delta$ (i.e., doping) and arbitrary $q$
and select the solution with $q$ minimizing the free energy.

{\it Results.} The results of our calculations are shown in
Figs.~\ref{fig:comm} - \ref{fig:incomm3}.
We find that the system behavior depends on the ratio $T_s/T_c >1$.
When this ratio is close to unity,
the system only develops a commensurate SDW order
with $m_{q=0}= m$ (see Fig.\ref{fig:comm}).
The SDW and SC transitions at $T_s (\delta)$ and $T_c$
(which is independent of $\delta$, see Eq.~(\ref{eq:scSCt}) with $m_\vq=0$)
are of second order.
Below tricritical point at which $T_s (\delta) = T_c$,
the transition between the states $(m \neq 0, \Delta =0)$
and $(\Delta \neq 0, m=0)$, is first order,
and there is no region where $m$ and $\Delta$ co-exist.

The first-order transition at $T=0$ can be understood analytically.
Setting $q=0$ and subtracting Eq. (\ref{eq:scSCt}) from (\ref{eq:scSDWt}),
we obtain for small $\delta$:
\be
\ln\frac{T_s}{T_c} = \frac{\delta^2}{m^2 + \Delta^2} \,.
\label{eq:tu1}
\ee
Setting $\Delta^2=0$ yields a linearized SC gap equations~(\ref{eq:scboth}).
We see that in the presence of a nonzero $m_0=m(T=0)$,
$\Delta$ first appears at $\delta^2_\Delta = m^2_0 \ln (T_s/T_c)$.
Similarly, for nonzero $\Delta_0$, the SDW order grows from 
$\delta^2_m = \Delta^2_0 \ln (T_s/T_c)$.
Their ratio, $\delta_\Delta/\delta_m = m_0/\Delta_0 = T_s/T_c > 1$,
implying that $\Delta$ nucleates in the
SDW phase at a higher doping whereas
$m$ develops in SC state at lower $\delta$ (see Fig.\ref{fig:comm}b).
This contradicts the very fact that SDW state is stable
at smaller dopings than the SC state.
The solution with $\Delta,m \neq 0$ then
grows in ``wrong'' direction of $\delta$, and we explicitly verified that
 is has a higher energy than pure states and therefore is 
unstable, see Fig. \ref{fig:comm}c.
As both $q=0$ SDW and SC gaps cover the entire FS, the absence of the
state where the two co-exist implies that
fully gapped SC and SDW orders cannot co-exist, and only one of these
two states is present at a given $\delta$. First order transition between the
SDW and SC states occurs at $\delta = \delta_{cr}$,
when their free energies coincide.
This happens when $-m_0^2/2+\delta^2_{cr} = -\Delta_0^2/2$, hence
\be
\delta^2_{cr} = \frac{m_0^2}{2} \left[ 1 - \left({T_c\over T_s}\right)^2\right] \,,
\ee
which is in between the two second order instability points
$\delta_m$ and $\delta_\Delta$.

Situation changes when $T_s/T_c$ gets larger, and there appears a wider range
of dopings where $T_s (\delta) > T_c$.  If only commensurate magnetic
order SDW$_0$ was possible, magnetic transition would become
first-order below a certain $T^*_s=0.56 T_s$, which at large enough $T_s/T_c$
becomes greater than $T_c$. In reality, the system avoids a first-order
transition and extends the region of magnetic order by forming
an incommensurate SDW$_q$ state below $T^*_s$
(see Figs.~\ref{fig:incomm5}, \ref{fig:incomm3}).\cite{rice70,kulikov84,cvetkovic08}
The transition from the normal state to SDW$_q$ state is second order,
the subsequent transition to the commensurate SDW$_0$ state is first order.
Once $m_q$ is developed, the actual solution is more complex
and includes harmonics with multiple $q$ (Ref. \onlinecite{kulikov84}),
but we ignore this for now.

Our main result is the discovery of a new phase below $T_c$, in which
incommensurate SDW$_q$ order co-exists with $s^+$ SC order.
Physically, the key reason for appearance of such phase is that
incommensurate SDW$_q$ order does not gap the excitations on entire FS --
the system remains a metal albeit with a modified, smaller
FS.\cite{cvetkovic08,mazin08a,mazin08b}  Once the Fermi surface survives, an
attractive pairing interaction gives rise to SC below $T_c$.
Alternatively speaking, for SDW$_q$ order,
some parts of the FS become unaccessible to magnetic
`pairing', and SC order takes advantage of this, 
c.f. Refs.~\onlinecite{machida81,machida87}.

At large enough $T_s/T_c$, the co-existence phase extends to $T=0$
(see Fig.~\ref{fig:incomm5} for $T_s/T_c =5$).
In Fig.~\ref{fig:incomm5}b we show order parameters at $T = T_c/2$.
The transition from a commensurate SDW state into a mixed state
is first order with  both  $\Delta$ and the amount of
incommensurateness $q$ jumping to finite values.
The transition  from a SDW$_q$ state into a mixed state, 
as well as the transition from a mixed
state into a pure SC state are of second order
with $m_q$ gradually vanishing.
Fig.~\ref{fig:incomm5}c shows the corresponding free energies of all states.
Comparing it with Fig.~\ref{fig:comm}c we see that now the
SC state becomes unstable at a higher $\delta$ and
the  mixed state now has lower energy than
pure SC or SDW states.

The behavior at somewhat smaller $T_s/T_c$ is intermediate between those in
Figs. \ref{fig:comm} and ~\ref{fig:incomm5}.  In figure~\ref{fig:incomm3} we
show the phase diagram for intermediate  $T_s/T_c = 3$. We still have
$T^*>T_c$, and the mixed phase still exists, but it now appears only as a small
pocket near the point where $T_s (\delta) =T_c$. At smaller $T$, the system
shows the same behavior as at $q=0$, i.e., a first-order transition between
commensurate SDW and SC states.

To conclude, in this paper we considered SDW and $s^+$ SC instabilities in a
mean-field approximation for a two-band model for the pnictides.  We assumed that
there are ``attractive'' interactions in both SDW and extended s-wave SC
channels, leading to the nesting-driven
SDW magnetism at $T_s (\delta)$, decreasing with $\delta$ controlled by
doping or pressure,
and to  $s^+$ SC at $T_c$. At zero doping, $T_s (0) = T_s >T_c$,
but $T_c$ becomes the first instability at large dopings.
The issue we consider is how SDW ordered
state transforms into a superconducting $s^+$ state.
We found that the transition is
first order, with no intermediate mixed phase, if $T_s$ and $T_c$  are close
enough so that SDW order above $T_c$ is commensurate.
At larger $T_s/T_c$, the system develops
an incommensurate SDW$_q$ order above $T_c$.
This incommensurate SDW
order does not gap the whole FS and allows for a
co-existence of magnetism and superconductivity. We found
that the mixed phase first appears in a small pocket near $T_s (\delta)
= T_c$, but extends as $T_s/T_c$ increases eventually reaching down to
$T=0$. At even larger $T_s/T_c$, zero temperature phases include
a commensurate SDW phase, an $s^+$ SC phase,
and an intermediate mixed phase where
$s^+$ SC co-exists with an incommensurate SDW order,
but there is no purely incommensurate $SDW$ at $T=0$.
The transition from the mixed phase to a commensurate SDW phase is first
order, and to the SC phase is second order.

Finally,  in this paper, we only considered a mixed state with a
uniform SC order (zero total momentum of a pair).
In principle, an inhomogeneous SC state (a true FFLO state) is possible because
incommensurate SDW$_q$ order breaks the symmetry between FS points with $\vk$
and $-\vk$. This, however, should not change the phase diagram   as
a non-uniform SC state  may only exist at large enough $m_q$,
i.e., near  a first order transition into a commensurate SDW phase,
possibly extending a mixed state into SDW$_0$ region.
For large $m_q$, an approximation by a single $q$ is not sufficient.
A more sophisticated analysis in this region is called for.

We acknowledge with thanks useful discussions with D.~Maslov, V.~Cvetkovic, I.~Eremin,
I.~Mazin, and Z.~Tesanovic. This work  was supported by NSF-DMR 0604406 (A.V.Ch).


\end{document}